\RequirePackage{filecontents}

\RequirePackage{snapshot}
\documentclass[aps,pra,reprint,showpacs,secnumarabic]{preprint}

\makeatletter
\def\lrbox#1{%
  \edef\reserved@a{%
    \endgroup
    \global\setbox#1\hbox{%
      \begingroup\aftergroup}%
    \def\noexpand\@currenvir{\@currenvir}%
    \def\noexpand\@currenvline{\on@line}}%
  \reserved@a
  \@endpefalse
  \color@setgroup
  \ignorespaces}
\def\endlrbox{\unskip\color@endgroup}
\makeatother

\renewcommand\documentclass[2][]{}
\newcommand{\changefont}[3]{}
\newcommand{\journal}[1]{}
\newbox\mybox
\newenvironment{frontmatter}{}{\maketitle}
\def\sep{\unskip, }%
\let\ead=\email
\let\PACS=\pacs
\newenvironment{keyword}{\begin{lrbox}{\mybox}\footnotesize}{\end{lrbox}\keywords{\usebox{\mybox}}}
\AtBeginDocument{%
  \IfFileExists{date.tex}{%
    \input{date.tex}

  }
}
\newcommand{\secreferences}{}
\pdfoutput=1
\documentclass[fleqn,final,5p,times,twocolumn]{elsarticle}

\usepackage[T1]{fontenc}
\usepackage{microtype}
\usepackage{graphicx}
\graphicspath{{figures/}}
\usepackage{amsmath}
\usepackage{bm}
\usepackage[pdftitle={Krylov subspace methods for the Dirac equation},
pdfauthor={Randolf Beerwerth and Heiko Bauke},
pdfcreator={},
pdfsubject={},
pdfproducer={},
pdfkeywords={Dirac equation; Krylov subspace methods; Lanczos algorithm},
hyperfootnotes=false,naturalnames]{hyperref}
\usepackage{algorithmic}
\usepackage{booktabs}
\usepackage[usenames,dvipsnames]{color}
\usepackage{siunitx}
\usepackage{mdwlist}
\DeclareSIUnit[number-unit-product=\,]{\au}{au}
\usepackage{braket}

\renewcommand*{\vec}[1]{\boldsymbol{#1}}
\newcommand*{\mat}[1]{\boldsymbol{\mathrm{#1}}}

\newcommand*{\I}{\mathrm{i}}
\newcommand*{\D}{\mathrm{d}}
\newcommand*{\e}{\mathrm{e}}

\newcommand*{\abs}[1]{\left\lvert#1\right\rvert}
\newcommand*{\trans}[1]{#1^{\mathsf{T}}}
\DeclareMathOperator{\spn}{span}
\DeclareMathOperator{\sgn}{sgn}
\DeclareUrlCommand\code{\urlstyle{tt}}

\providecommand{\secreferences}{\section*{References}}

\begin{document}

\begin{frontmatter}
  \title{Krylov subspace methods for the Dirac equation}
  \author{Randolf Beerwerth}
  \ead{randolf.beerwerth@mpi-hd.mpg.de}
  \author{Heiko Bauke}
  \ead{heiko.bauke@mpi-hd.mpg.de}
  \address{Max-Planck-Institut f{\"u}r Kernphysik, Saupfercheckweg~1,
    69117~Heidelberg, Germany}
  \begin{abstract}
    The Lanczos algorithm is evaluated for solving the time-independent
    as well as the time-dependent Dirac equation with arbitrary
    electromagnetic fields.  We demonstrate that the Lanczos algorithm
    can yield very precise eigenenergies and allows very precise time
    propagation of relativistic wave packets. The unboundedness of the
    Dirac Hamiltonian does not hinder the applicability of the Lanczos
    algorithm.  As the Lanczos algorithm requires only matrix-vector
    products and inner products, which both can be efficiently
    parallelized, it is an ideal method for large-scale calculations.
    The excellent parallelization capabilities are demonstrated by a
    parallel implementation of the Dirac Lanczos propagator utilizing
    the Message Passing Interface standard.
  \end{abstract}
\end{frontmatter}

\noindent
{\bf PROGRAM SUMMARY}\\
\begin{small}
\noindent
{\em Program title:} Dirac\_Laczos \\[0.5ex]
{\em Catalogue identifier:} AEUY\_v1\_0 \\[0.5ex]
{\em Program summary URL:}\\ \url{http://cpc.cs.qub.ac.uk/summaries/AEUY_v1_0.html} \\[0.5ex]
{\em Program obtainable from:} CPC Program Library, Queen's University, Belfast, N. Ireland\\[0.5ex]
{\em Licensing provisions:} Standard CPC licence, 
\url{http://cpc.cs.qub.ac.uk/licence/licence.html}\\[0.5ex]
{\em No. of lines in distributed program, including test data, etc.:} 526\,428\\[0.5ex]
{\em No. of bytes in distributed program, including test data, etc.:} 2\,181\,729\\[0.5ex]
{\em Distribution format:} tar.gz\\[0.5ex]
{\em Programming language:} C++11\\[0.5ex]
{\em Computer:} multi-core systems or cluster computers \\[0.5ex]
{\em Operating system:} any\\[0.5ex]
{\em Has the code been vectorized or parallelized?:} parallelized using MPI\\[0.5ex]
{\em RAM:} typically 10 megabyte to 1 gigabyte depending on the chosen
probelem size\\[0.5ex]
{\em Classification:} 2.7\\[0.5ex]
{\em External routines/libraries:} Boost [1], LAPACK [2]\\[0.5ex]
{\em Nature of problem:} solving the time-dependent Dirac equation in
two spatial dimensions \\[0.5ex]
{\em Solution method:} Lanczos propagator\\[0.5ex]
{\em Running time:} depending on the problem size and computer hardware
typically several minutes to several days\\[2.5ex]
References:
\begin{itemize*}
\item[{[1]}] Boost C++ Libraries, \url{http://www.boost.org}
\item[{[2]}] LAPACK--Linear Algebra PACKage, \url{http://www.netlib.org/lapack/}
\end{itemize*}
\end{small}
\pagebreak[1]

\section{Introduction}
\label{sec:intro}

The Dirac equation is the fundamental equation of motion for describing
the quantum evolution of a charged spin one-half particle in a Lorentz
invariant manner.  It predicts the existence of antimatter and finds its
application not only in the growing field of light-matter interactions
at relativistic intensities
\cite{Salamin:Hu:etal:2006:Relativistic_high-power,Ehlotzky:etal:2009:Fundamental_processes,DiPiazza:etal:2012:extremely_high-intensity}
but also in condensed matter theory of graphene
\cite{Brey:Fertig:2006:Electronic_states} and in relativistic quantum
information
\cite{Peres:Terno:2004:Quantum_information,Friis:Bertlmann:etal:2010:Relativistic_entanglement,Saldanha:Vedral:2012:Physical_interpretation}.
Deducing analytical solutions of this equation, however, poses a major
problem. Analytical methods for determining solutions of the Dirac
equation usually require physical setups with a high degree of symmetry
\cite{Bagrov:Gitman:1990:Exact_Solutions,Gross:1999:Relativistic_QM,Strange:1998:Relativistic_QM,Schwabl:2008:Advanced_QM,Wachter:2009:Relativistic_QM}.
Thus, approximations or numerical methods have to be applied.  In the
past, various numerical schemes have been developed to solve the
time-dependent Dirac equation numerically, including Fourier split
operator approaches
\cite{Braun:etal:1999:Numerical_approach,Mocken:Keitel:2004:Quantum_dynamics,Mocken:Keitel:2008:FFT-split-operator,Bauke:Keitel:2011:Accelerating},
real space split operator methods based on the method of characteristics
\cite{Fillion-Gourdeau:Lorin:etal:2012:Numerical_solution_time-dependent,Fillion-Gourdeau:Lorin:etal:2014:split-step_numerical},
finite differences \cite{Hammer:Potz:2014:Staggered_grid}, finite
elements \cite{Muller:etal:1998:Finite_element_Dirac}, or employing
either spherical harmonics or plane waves as basis functions and
integrating the resulting ordinary differential equations
\cite{Selst:Lindroth:etal:2009:Solution_of,Ahrens:etal:2013:Kapitza-Dirac}.
Complementary to the mentioned numerical methods quantum simulations of
the Dirac equation
\cite{Gerritsma:Kirchmair:etal:2010:Quantum_simulation} are also an
active field of current research.

State of the art large scale calculations of the time-dependent
Schr{\"o}dinger equation
\cite{Schneider:etal:2006:RSP-FEDVR,Guan:Noble:etal:2009:ALTDSE_Arnoldi-Lanczos},
which is the nonrelativistic limit of the Dirac equation, often utilize
Krylov subspace methods and in particular the Lanczos algorithm
\cite{Lanczos:1950:iteration_method,Golub:Van_Loan:1996:Matrix_Computations,Saad:2011:Large_Eigenvalue_Problems}.
Although the Lanczos algorithm was known since the 1950s, this approach
gained popularity not until the 1980s when it was applied to
time-independent \cite{Haydock:1980:recursive_solution} and
time-dependent
\cite{Park:Light:1986:Unitary_quantum,Leforestier:etal:1991:propagation_schemes}
problems of nonrelativistic quantum mechanics.  Krylov subspace methods
have the virtue that their applicability does not depend on the manner
how time and space are discretized.  Furthermore, they are rather easy
to parallelize and therefore suitable for today's prevalent parallel
hardware architectures like compute clusters
\cite{Bauke:Mertens:2005:ClusterComputing} or high-performance graphics
cards \cite{Wilt:2013:CUDA_Handbook}.  As the Dirac equation and the
Schr{\"o}dinger equation share the same Hermitian structure, it appears
natural to apply Krylov subspace methods also to relativistic quantum
mechanical problems.  Thus, it is the purpose of this article to
evaluate the performance of the Lanczos algorithm when it is applied to
the relativistic Dirac equation.

This paper is organized as follows. In order to make the paper
self-contained and to introduce some notations we characterize the
Lanczos algorithm in Section~\ref{sec:Lanczos} and show how it can help
to solve Hermitian eigenvalue problems. In Section~\ref{sec:Dirac} we
will briefly cover the Dirac equation. Furthermore, we will explain in
Section~\ref{sec:matrix_exponential} how the time evolution operator can
be approximated using the Lanczos algorithm. Numerical results are
presented in Section~\ref{sec:Eigen} for the relativistic eigenproblem
and in Section~\ref{sec:time} for the time-dependent Dirac equation.
Finally, we will present our parallel implementation of the Lanczos
Dirac propagator and show some benchmark results in
Section~\ref{sec:implementation}.

\section{The Lanczos algorithm}
\label{sec:Lanczos}

Before we will briefly summarize the Lanczos algorithm and its
properties we have to introduce the notion of a Krylov subspace.  A
Krylov subspace of dimension $k$ generated by an $N \times N$ matrix
$\mat A$ is spanned by the successive powers of $\mat A$ applied to some
given vector $\vec b$
\begin{equation}
  \mathcal{K}_k(\mat A, \vec b) =
  \spn \left\{ 
    \vec b, \mat A \vec b, \mat A^2 \vec b, \dots, \mat A^{k-1} \vec b
  \right\}\,,\quad 1\leq k \leq N \,.
\end{equation}
For every Hermitian matrix $\mat A$ there is a unitary transformation
that turns the matrix into tridiagonal form, that is,
\begin{equation}
  \label{eq:krylov_unitary_transform}
  \mat Q^\dagger \mat A \mat Q = \mat T = 
  \begin{pmatrix}
    \alpha_1 &  \beta_1 \\
    \beta_1 & \alpha_2 &  \beta_2 \\
    & \ddots & \ddots & \ddots \\
    & & \beta_{N-2} & \alpha_{N-1} &  \beta_{N-1} \\
    & & & \beta_{N-1} & \alpha_N 
  \end{pmatrix}\,.
\end{equation}
The unitary transformation matrix $\mat Q$ and the tridiagonal matrix
$\mat T$ can be determined via the Lanczos algorithm.  The columns of
the matrix $\mat Q$ are conveniently labeled $\vec q_i$ and are called
the Lanczos vectors. The vectors $\vec q_1$ to $\vec q_k$ form an
orthonormal basis of the Krylov subspace $\mathcal{K}_k(\mat A, \vec
b)$.  The Lanczos algorithm is an iterative procedure that is based on
the two equations
\begin{subequations}
  \label{eq:lanczos_base}
  \begin{align}
    \label{eq:lanczos_gram_schmidt_alpha}
    \alpha_i & = 
    \vec q_i \cdot  \mat A \vec q_i \,, \\
    \label{eq:lanczos_gram_schmidt}
    \beta_i \vec q_{i+1} & = 
    \mat A \vec q_i - \alpha_i \vec q_i - \beta_{i-1} \vec q_{i-1} \,. 
  \end{align}
\end{subequations}
These equations correspond to a classical Gram-Schmidt
orthogonalization.  One can show that due to the Hermiticity of $\mat
A$, the newly constructed vector $\vec q_{i+1}$ is automatically
orthogonal to all previous Lanczos vectors, except the last
two. Therefore, the classical Gram-Schmidt orthogonalization reduces to
the subtraction of the contributions from two previous vectors, shown in
Eq.~\eqref{eq:lanczos_gram_schmidt}.

\begin{figure}\noindent
  \begin{algorithmic}	
    \STATE $\vec q_1 = {\vec b}/{\lVert \vec b\rVert_2 }$
    \STATE $\vec z = \mat A \vec q_1$
    \FOR{$i=1$ \TO $k-1$}
    \STATE $\alpha_i = \vec q_i \cdot \vec z$
    \STATE $\vec z = \vec z - \alpha_i \vec q_i$    
    \STATE $\beta_i = \lVert \vec z \rVert_2$    
    \STATE $\vec q_{i+1} = {\vec z}/{\beta_{i}}$
    \STATE $\vec z = \mat A \vec q_{i+1} - \beta_i \vec q_i$
    \ENDFOR
    \STATE $\alpha_k = \vec q_k \cdot \vec z$    
  \end{algorithmic}
  \caption{The Lanczos algorithm calculates the orthonormal basis $\vec
    q_1$, $\vec q_2$, \dots,$\vec q_k$ of the Krylov subspace
    $\mathcal{K}_k(\mat A, \vec b)$ of the $N\times N$ matrix $\mat A$.
    For $N=k$ the matrix $\mat Q=[\vec q_1,\dots,\vec q_N]$ and the
    tridiagonal matrix $\mat T$ are defined in
    \eqref{eq:krylov_unitary_transform} such that the matrix $\mat A$
    factorizes as $\mat A = \mat Q \mat T \mat Q^\dagger$.}
  \label{fig:Lanczos}
\end{figure}

Due to the well-known stability issues of the classical Gram-Schmidt
algorithm, Paige \cite{Paige:1976:Error_Analysis} suggested to replace
\eqref{eq:lanczos_base} by a modified Gram-Schmidt orthogonalization
against the last two previous vectors.  With the modified Gram-Schmidt
orthogonalization Eq.~\eqref{eq:lanczos_gram_schmidt_alpha} becomes
\begin{equation}
  \alpha_i = 
  \vec q_i \cdot \left(\mat A \vec q_i - \beta_{i-1} \vec q_{i-1}\right) \,,
\end{equation}
which is in exact arithmetic equivalent to
\eqref{eq:lanczos_gram_schmidt_alpha} because $\vec q_i$ and $\vec
q_{i-1}$ are orthogonal.  This leads to the Lanczos algorithm as shown
in Fig.~\ref{fig:Lanczos}. Using an arbitrary nonzero starting vector
$\vec b$ it calculates the columns of the matrix $\mat Q$ iteratively
such that \eqref{eq:krylov_unitary_transform} is fulfilled.  After each
iteration with $1<k<N$ the relation
\begin{equation}
  \label{eq:Lanczos_relation}
  \mat A \mat Q^{(k)} = 
  \mat Q^{(k)} \mat T^{(k)} + 
  [\underbrace{\vec 0, \dots, \vec 0}_{k-1 \text{ times}}, \beta_k \vec q_{k+1}] \,,
\end{equation}
which is frequently called the Lanczos relation, holds, where $\mat
Q^{(k)}$ is formed by the first $k$ column vectors $\vec q_i$ and $\mat
T^{(k)}$ is the symmetric tridiagonal matrix formed by $\alpha_1$ to
$\alpha_k$ and $\beta_1$ to $\beta_{k-1}$. Because the Lanczos algorithm
in Fig.~\ref{fig:Lanczos} performs only matrix-vector products (and
inner products) it is sufficient if the action of the matrix $\mat A$ on
some vector can be computed.  It is not necessary to store the matrix
elements of $\mat A$ explicitly, which is a major advantage of the
Lanczos algorithm.

Since the transformation $\mat Q$ is unitary, the matrices $\mat A$ and
$\mat T$ are similar and therefore they have the same set of eigenvalues
and the eigenvectors $\vec a_i$ of $\mat A$ are related to the
eigenvectors $\vec t_i$ of $\mat T$ via
\begin{equation}
  \vec a_i = \mat Q\vec t_i \,.
\end{equation}
In many applications it is sufficient to know some eigenvalues and
eigenvectors of $\mat A$.  Because the matrix $\mat T^{(k)}$ is the
representation of $\mat A$ in the Krylov subspace $\mathcal{K}_k(\mat A,
\vec b)$ and as a consequence of the Lanczos relation
\eqref{eq:Lanczos_relation} the approximation
\begin{equation}
  \label{eq:A_approx} 
  \mat A \approx \mat Q^{(k)} \mat T^{(k)} \mat Q^{{(k)}^\dagger} 
\end{equation}
holds.  Thus, some of the eigenvalues $\lambda_i$ of $\mat A$ may be
approximated by the eigenvalues $\lambda_i^{\left(k\right)}$ of $\mat
T^{(k)}$ with $k \ll N$, that is when the Lanczos iteration is stopped
after $k$ iterations.  Approximate eigenvectors of $\mat A$ can be
obtained by
\begin{equation}
  \label{eq:eigenvec_approx}
  \vec a_i \approx \vec a_i^{(k)} = \mat Q^{(k)} \vec t_i^{(k)} \,.
\end{equation}

The error of the approximate eigenvalues is bounded by
\cite{Demmel:1997:Applied_Numerical}
\begin{equation}
  \label{eq:error_est}
  \Delta \lambda_i^{(k)} =
  \min\limits _{j}  \big| \lambda_i^{(k)} - \lambda_j \big| \leq 
  \abs{\beta_k t_i^k} \,,
\end{equation}
where $t_i^k$ denotes the $k$th (the last) component of the $i$th
eigenvector of the matrix $\mat T^{(k)}$.  Furthermore, $\abs{\beta_k
  t_i^k}$ equals the residual of the $i$th eigenvector, that is,
\begin{equation}
  \left\lVert \mat A \vec a_i^{(k)} - \lambda_i^{(k)} \vec a_i^{(k)} \right\rVert_2 = 
  \abs{\beta_k t_i^k} \,,
\end{equation}
where $\lVert\cdot\rVert_2$ denotes the Euclidean vector norm.  The
error estimate \eqref{eq:error_est} allows one to monitor the error
after each iteration and to stop as soon as a sufficient accuracy has
been reached.

A notable deficiency of the simple Lanczos algorithm as presented here
is its numerical instability.  In exact arithmetic the matrix $\mat Q$
is unitary. This property, however, is lost in floating point arithmetic
when rounding errors occur. Paige's theorem
\cite{Demmel:1997:Applied_Numerical} shows that the Lanczos vector $\vec
q_i$ looses its orthogonality with respect to the other Lanczos vectors
as the corresponding eigenvalue converges.  Thus, convergence comes at
the price of loss of orthogonality.  The numerical stability can,
however, be increased by applying reorthogonalization.  This means, the
orthogonalization against the two previous Lanczos vectors is replaced
by an orthogonalization against all previous Lanczos vectors using
either the classical Gram-Schmidt algorithm or the modified Gram-Schmidt
algorithm. This is still not unconditionally stable.  Unconditional
stability is ensured by applying orthogonalization twice
\cite{Parlett:1998:Symmetric_Eigenproblem}. This extension of the
Lanczos algorithm is usually called the Lanczos algorithm with full
reorthogonalization. In the following, it will be used to demonstrate
the convergence behavior and to demonstrate that degenerate states are
found correctly.  An alternative to the very expensive full
reorthogonalization is partial reorthogonalization
\cite{Simon:1984:Lanczos_Partial_Reorthogonalization}.  All
reorthogonalization approaches, however, have the common disadvantage
that they require all Lanczos vectors to be stored. If no
reorthogonalization is applied, storage of three vectors is sufficient
to determine the approximate eigenvalues.  If, however, also the
approximate eigenvectors \eqref{eq:eigenvec_approx} are required then
the whole matrix $\mat Q^{(k)}$ has to be stored.  Alternatively one can
run the Lanczos algorithm a second time to calculate the approximate
eigenvectors $\vec a_i^{(k)}$ after the vectors $\vec t_i^{(k)}$ have
been determined.  The application of the matrix $\mat Q^{(k)}$ to the
vector $\vec t_i^{(k)}$ is calculated while performing the second
Lanczos iteration.

\section{The Dirac equation}
\label{sec:Dirac}

The Dirac equation 
\begin{equation}
  \label{eq:Dirac}
  \I \hbar \frac{\partial \Psi(\vec x, t)}{\partial t} = 
  \mat{\hat H}(t) \Psi(\vec x, t)
\end{equation}
describes a relativistic spin one-half particle with rest mass $m$ and
charge $q$ moving in the electromagnetic potentials $\vec A\left(\vec x,
  t\right)$ and $\phi\left(\vec x, t\right)$.  The in general
time-dependent Hamiltonian $\mat{\hat H}(t)$ is in $d$ spatial
dimensions given by
\begin{equation}
  \mat{\hat H}(t) = 
  c \sum_{i = 1}^d \alpha_i \left(
    \hat p_i - q A_i\left(\vec x, t\right)
  \right) + \beta m c^2 + q \phi\left(\vec x, t\right) \,,
\end{equation}
where $\hat{p}_i=-\I\hbar\partial_{x_i}$ and $A_i\left(\vec x, t\right)$
denote the components of the canonical momentum operator and the vector
potential, respectively, and $\alpha_i$ and $\beta$ are the Dirac
matrices. These matrices obey the Dirac algebra
\begin{subequations}
  \begin{align}
    \alpha_i^2 =  \beta^2 & = 1 \,,\\
    \alpha_i  \beta +  \beta  \alpha_i & = 0 \,,\\
    \alpha_i  \alpha_j +  \alpha_j  \alpha_i & = 2\delta_{i,j} \,,
  \end{align}\label{eq:dirac_matrices_conditions}%
\end{subequations}
for $i, j = 1, \dots, d$.  In one dimension the Dirac matrices are
given by $2\times2$ matrices. The choice $\alpha_1=\sigma_1$ and
$\beta=\sigma_3$ with $\sigma_1$, $\sigma_2$, and $\sigma_3$ denoting
the three Pauli matrices is the standard representation for
one-dimensional systems. In three space dimensions, however, $4\times4$
matrices are required with the standard representation
\begin{align}
  \label{eq:dirac_matices_4x4}
  \alpha_i & =
  \begin{pmatrix}
    0 & \sigma_i \\ \sigma_i & 0
  \end{pmatrix}\,,
  &
  \beta & =
  \begin{pmatrix}
    1 & 0 \\ 0 & -1
  \end{pmatrix}\,.
\end{align}
The $2\times2$ Pauli matrices also fulfill the Dirac algebra in two
dimensions with $\alpha_1=\sigma_1$, $\alpha_2=\sigma_2$, and
$\beta=\sigma_3$.  The resulting two-dimensional Dirac equation does,
however, not include the spin degree of freedom.  To incorporate the
electron spin into a two-dimensional Dirac equation the $4\times4$
matrices \eqref{eq:dirac_matices_4x4} have to be employed. Because the
Dirac Hamiltonian is a matrix operator Dirac wave functions have two or
four complex components.  As in the nonrelativistic case the eigenvalue
equation resulting from the time-independent problem reads
\begin{equation}
  \label{eq:Dirac_eigen}
  \hat {\mat H} \Psi(\vec x) = E \Psi(\vec x) \, .
\end{equation}
The spectrum of a free particle consists of a positive and a negative
continuum $(-\infty, -m c^2] \cup [m c^2, \infty)$, which are usually
associated with particles and anti-particles.  

A formal solution of the time-dependent problem of the Dirac equation
\eqref{eq:Dirac} can be given using the time-ordering operator. For
numerical calculations we neglect time ordering and approximate the time
evolution operator $U\left(t, t + \Delta t\right)$ in first order via
the first term of a Magnus expansion
\cite{Pechukas:1966:On_Exponential,Blanes:Casas:etal:2009:Magnus_expansion},
which results in
\begin{multline}
  \label{eq:time_evolution_approx}
  \Psi(\vec x, t + \Delta t) = 
  U\left(t, t + \Delta t\right)\Psi(\vec x, t) \approx \\
  \exp\left(
    -\frac{\I}{\hbar} \int_{t}^{t+\Delta t} \hat{\mat H}(t')\,\D t'
  \right) \Psi(\vec x, t)  
  + \mathcal{O}\left(\Delta t^3\right) \,.
\end{multline}
By discretization of the wave function $\Psi(\vec x, t)$ the Hamiltonian
$\hat{\mat H}(t)$ becomes a matrix.  Thus, the numerical propagation of
the wave function $\Psi(\vec x, t)$ by one time step $\Delta t$ requires
the calculation of a matrix exponential, followed by the application to
the state vector.  Similarly, the numerical solution of the
time-independent problem \eqref{eq:Dirac_eigen} involves the computation
of the eigenvalues and the eigenvectors of a matrix.

\section{Time propagation and matrix exponentials}
\label{sec:matrix_exponential}

The standard method for computing the exponential of a Hermitian matrix
is to calculate its eigendecomposition.  Then, the matrix exponential in
the space of the matrix' eigenvectors reduces to exponentials of the
eigenvalues.  Typical matrices that result from the discretization of
partial differential equations, however, are so large that full
diagonalization is not feasible.  Therefore, the matrix exponential is
calculated approximately in a Krylov subspace of dimension much smaller
than the dimension of the original matrix.

Let $\mat A$ now denote the Hermitian matrix that results from the
discretization of the operator $\int_{t}^{t+\Delta t} \hat{\mat
  H}(t')\,\D t'/\hbar$ and $\vec\psi(t)$ the corresponding discrete
representation of the wave function $\Psi(\vec x, t)$.  Utilizing the
approximation \eqref{eq:A_approx} and the relation
\eqref{eq:krylov_unitary_transform} the exponential of the antihermitian
matrix $-\I\mat A$ applied to the vector $\vec\psi$ can be calculated
approximately in the Krylov subspace $\mathcal{K}_k(\mat A,
\vec\psi(t))$ as \cite{Saad:1992:Analysis}
\begin{multline}
  \label{eq:matrix_exp_approx}
  \vec \psi(t+\Delta t) =
  \exp{\left(- \I \mat A\right)} \vec \psi(t) =
  \mat Q \exp\left(-\I \mat T\right) \mat Q^\dagger\vec \psi(t) \approx \\
  \mat Q^{(k)} \exp\left(-\I \mat T^{(k)}\right) {\mat
    Q^{(k)}}^\dagger\vec \psi(t) =
  \lVert\vec\psi(t)\rVert_2 \mat Q^{(k)} \exp\left(-\I \mat T^{(k)}\right) \vec e_1 \,,
\end{multline}
where $\vec e_1$ denotes the $k$-dimensional unit vector $\trans{(1, 0,
  0, \dots)}$.  The remaining matrix exponential of the matrix $\mat
T^{(k)}$ can be computed by performing a numerical eigendecomposition
using any standard method optimized for tridiagonal real symmetric
matrices. This diagonalization is not very expensive because the
dimension of $\mat T^{(k)}$ is chosen to be very small compared to the
dimension of $\mat A$.  Note that it is required not to include the
factor $-\I$ into the matrix $\mat A$ in order to ensure the Hermiticity
of $\mat A$ and thus the applicability of the Lanczos algorithm.  Note
that \eqref{eq:matrix_exp_approx} involves all Lanczos vectors $\vec
q_1$ to $\vec q_k$.  This means that these vectors have to be stored and
cannot be discarded when no longer needed in the Lanczos algorithm.  In
order to save memory one can run the Lanczos algorithm a second time
when the matrix $\mat Q^{(k)}$ is multiplied to the vector $\exp(-\I
\mat T^{(k)})\, \vec e_1$.

The error that is introduced by the approximation
\eqref{eq:matrix_exp_approx} may be estimated as
\cite{Saad:1992:Analysis}
\begin{equation}
  \label{eq:matrix_exponential_error}
  \lVert \Delta\vec\psi \rVert_2 \approx
  \beta_{k} \lVert \vec \psi(t) \rVert_2 \abs{\trans{\vec e_k} \exp\left(-\I\mat T^{(k)}\right) \vec e_1} \,,
\end{equation}
where $\Delta\vec\psi$ is the residual vector $\vec \psi(t+\Delta
t)-\exp(-\I\mat{A})\vec \psi(t)$.  This error estimate can be used to
monitor the accuracy of a calculation and to adjust the size of the time
step $\Delta t$ or the dimension of the Krylov subspace $k$ adaptively.
However, the number of Lanczos iterations should be kept on a moderate
level since the orthogonality of the Lanczos vectors may be lost.  When
orthogonality is lost, more Lanczos iterations will not further improve
the approximation \eqref{eq:matrix_exp_approx}.  The estimate
\eqref{eq:matrix_exponential_error} is especially useful because it can
be computed with very little extra cost.  It is, however, very loose and
tends to overestimate the actual error.  There are more precise but much
more expensive to compute estimates, which can be found, for example, in
\cite{Saad:1992:Analysis}. Furthermore, the total error in solving the
time-dependent Dirac equation depends also on the discretization of the
Hamiltonian, which is not included in the estimate
\eqref{eq:matrix_exponential_error}.

\begin{figure*}
  \centering
  \includegraphics[width=\linewidth]{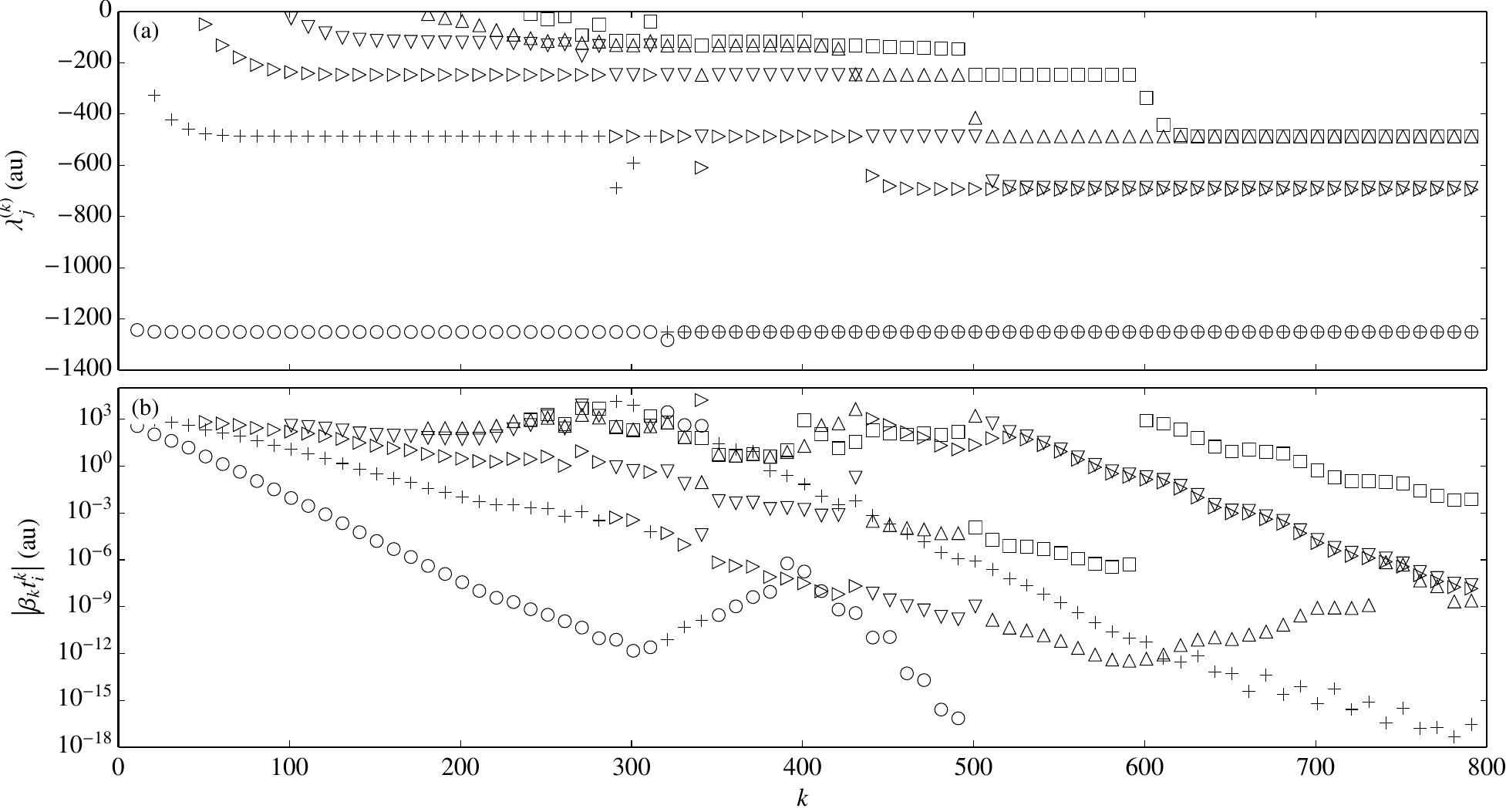}
  \caption{Iterative calculation of the eigenenergies of the
    two-dimensional Dirac Hamiltonian with the soft-core potential
    \eqref{eq:soft_core_pot} with $Z = 50$ via the Lanczos algorithm
    with full reorthogonalization. A basis set of $N = \num{64} $ basis
    functions per dimension was employed. Part (a) shows the eigenvalues
    (rest-mass energy $mc^2$ subtracted) of the first six bound states
    as a function of the iteration $k$, while (b) shows the error bound
    \eqref{eq:error_est}. For clarity data of only every 10th Lanczos
    iteration is presented in both plots.}
  \label{fig:soft_core_1}
\end{figure*}

\section{Time-independent problems}
\label{sec:Eigen}

Paige \cite{Paige:1972:Computational_Variants,Meurant:2006:Lanczos}
demonstrated that despite the loss of orthogonality the Lanczos
algorithm is a capable method to find few of the extreme eigenvalues of
large Hermitian matrices.  This makes it the ideal tool for computing
the bound states of nonrelativistic Hamiltonians, which lie at the lower
end of the energy spectrum.  In case of the Dirac equation, however, the
bound states appear inside the band gap in the middle of the spectrum.
Cullum and Willoughby
\cite{Cullum:Willoughby:1981:Computing_eigenvalues} pointed out that the
convergence rate crucially depends on the gap structure of the
eigenvalue spectrum. The $\lambda^{(k)}_j$ converge particularly fast to
eigenvalues of the matrix $\mat A$ that are well-separated from other
eigenvalues.  Thus, well-separated interior eigenvalues may converge
faster than clustered extreme eigenvalues.  In this section we will
demonstrate that the Lanczos algorithm is able to find also the
eigenvalues of bound states of Dirac Hamiltonians, which are neither
bounded from above nor bounded from below.  In fact, bound states in the
band gap converge much faster than states in the continuum part of the
spectrum.

\subsection{Soft-core potential in two dimensions}

As an illustrative example for the calculation of eigenstates we
consider the two-dimensional soft-core potential with
\begin{equation}
  \label{eq:soft_core_pot}
  q\phi\left(r\right) = 
  -\frac{3}{2}\frac{Ze^2}{4\pi\varepsilon_0\sqrt{r^2 + 3/Z^2}} \,.
\end{equation}
For the remainder of this article atomic units will be employed. In this
system of units the electron mass $m$, the elementary charge $e$, the
Bohr radius $a_0$ and the reduced Planck constant $\hbar$, and as a
consequence $1/(4\pi\varepsilon_0)$ are unity.  The speed of light
equals the inverse fine structure constant and energy is measured in
units of hartree, $\SI{1}{\au} = \SI{27.211385}{eV} = \SI{1}{E_h}$. The
magnetic field is measured in units of $\SI{1}{\au}=
\SI{2.35052e5}{T}$. The soft-core parameters in \eqref{eq:soft_core_pot}
are chosen such that the ground-state energy of this potential yields in
case of the nonrelativistic Schr{\"o}dinger equation $-Z^2/2$, which is the
same value as for the three-dimensional Coulomb potential with the same
atomic number $Z$.  The nonrelativistic normalized ground state of the
soft-core potential \eqref{eq:soft_core_pot} is
\cite{Clark:1997:Closed-form_solutions}
\begin{equation}
  \label{eq:SC_exact_gs}
  \Psi(r) = 
  \frac{2\e^{\!\!\sqrt{3}}\,Z\left(1+\sqrt{r^2Z^2+3}\right)}
  {\sqrt{\smash[b]{3\pi(10\sqrt 3 + 17)}}}
  \exp\left({-\sqrt{r^2Z^2+3}}\right)\,.
\end{equation}

The four-component Dirac wave functions are discretized using a
pseudospectral method.  For this purpose each component of the wave
function is expanded into a finite set of $N\times N$ two-dimensional
basis functions, which are defined via a tensor product of the first $N$
Hermite functions. During the calculation wave functions are represented
via function values at the collocation points (rather than by expansion
coefficients). The collocation points correspond to the roots of the
$N$th Hermite function.  Derivatives are then expressed by some dense
matrices. The choice of Hermite functions as the underlying basis
functions ensures that the wave function obeys the boundary condition
that $\Psi(\vec x, t)$ goes to zero as $|\vec x|\to\infty$.
Furthermore, the dense pseudospectral differentiation matrices allow for
an approximation of the differential operators that is much more
accurate than local finite difference schemes.  Note, that the discrete
variable representation
\cite{Light:Hamilton:etal:1985:Generalized_discrete,Szalay:1993:Discrete_variable}
of differential operators, which is often applied to solve the
nonrelativistic Schr{\"o}dinger equation, is also based on orthogonal
polynomials similarly to the employed pseudospectral method. Because the
Lanczos approach does not depend on a specific discretization the
discrete variable representation may also be applied to the Dirac
equation.

The convergence behavior of the Lanczos algorithm for the
two-dimensional Dirac equation with a soft-core potential with $Z=50$ is
shown in Fig.~\ref{fig:soft_core_1}.  Here a basis set of $N=64$ basis
functions per dimension was used, which corresponds to a Hamiltonian
matrix of size $4\times64^2=16384$.  A Gaussian wave packet of width
$\sim 1/Z$ was used as an initial starting vector for the Lanczos
algorithm. As the soft-core potential goes to zero for $r\to\infty$
bound states must lie in the range $(0, mc^2)$ and can therefore be
identified easily.  Figure~\ref{fig:soft_core_1}(a) shows the
eigenvalues of the six bound states with the lowest energy in $(0,
mc^2)$ as a function of the number of iterations $k$.  One can see that
the ground state converges very fast, excited states, however, follow
significantly later.  Note that the bound states do not appear in their
energetic order. The second spin state of the degenerate ground state,
for example, needs about 300 iterations to appear for the first time.
As a consequence of the Cauchy interlace theorem
\cite{Demmel:1997:Applied_Numerical,Hwang:2004:Cauchys_Interlace} it
appears initially between two already converged eigenvalues and then
converges to the correct value.
Figure~\ref{fig:soft_core_1}(b) shows for each eigenvalue of
Fig.~\ref{fig:soft_core_1}(a) the corresponding error estimate
\eqref{eq:error_est}.  After about 600 iterations both ground state
energies are converged to machine precision.
Figure~\ref{fig:soft_core_1} also illustrates that seemingly converged
eigenvalues may cross over to another value, as for example at
$k\approx300$, where the second ground state of the soft-core potential
appears.

\begin{table}\small
  \caption{Eigenenergies (minus the rest-mass energy $mc^2$) of the
    two-dimensional soft-core potential
    \eqref{eq:soft_core_pot}. Results that are obtained using the
    Lanczos algorithm with 1000 iterations are compared to numerical
    solutions calculated by a Fourier-spectral method. The same
    parameters as in Fig.~\ref{fig:soft_core_1} were used.}
  \label{tab:soft_core_eigenstates_Z_50}
  \vspace*{1ex}\centering
  \begin{tabular}{@{}lrr@{}}
    \toprule
    & \multicolumn{2}{c@{}}{energy in \si{\au}} \\
    \cmidrule(l){2-3}
    state number & \multicolumn{1}{c}{Lanczos} & \multicolumn{1}{c@{}}{Fourier} \\
    \cmidrule(r){1-1}\cmidrule(lr){2-2}\cmidrule(l){3-3}
    1 & $-1250.55965$ & $-1250.559$ \\
    2 & $-695.15042$ & $-695.151$ \\
    3 & $-688.06828$ & $-688.067$ \\
    4 & $-487.52777$ & $-487.529$ \\
    5 & $-380.4172\phantom{0}$ & $-380.425$ \\
    6 & $-376.4623\phantom{0}$ & $-376.471$ \\
    7 & $-320.86\phantom{000}$ & $-320.633$ \\
    8 & $-318.84\phantom{000}$ & $-318.601$ \\
    9 & $-247.66\phantom{000}$ & $-251.149$ \\
    \bottomrule
  \end{tabular}
\end{table}
\begin{table}\small
  \caption{Ground state energies (minus the rest-mass energy $mc^2$) of the
    two-dimensional soft-core potential \eqref{eq:soft_core_pot} for
    different values of the atomic number $Z$. The results obtained with
    the Lanczos algorithm using 1000 iterations are compared to
    perturbation-theory results. In addition the exact ground state
    energy for the corresponding nonrelativistic Schr{\"o}dinger equation is given.} 
  \label{tab:soft_core_groundstates}
  \vspace*{1ex}\centering
  \begin{tabular}{@{}rrrr@{}}
    \toprule
    & \multicolumn{3}{c@{}}{energy in \si{\au}} \\
    \cmidrule(l){2-4}
    \multicolumn{1}{c}{$Z$} & \multicolumn{1}{c}{Lanczos} &
    \multicolumn{1}{c}{perturbation theory} & \multicolumn{1}{c@{}}{nonrelativistic} \\
    \cmidrule(r){1-1}\cmidrule(lr){2-2}\cmidrule(lr){3-3}\cmidrule(l){4-4}
    1 & $-0.500000089$ & $-0.500000090$ & $-0.5$ \\
    2 & $-2.00000144\phantom{0}$ & $-2.000001441$ & $-2.0$\\
    3 & $-4.50000728\phantom{0}$ & $-4.500007297$ & $-4.5$\\
    5 & $-12.5000562\phantom{00}$ & $-12.50005638$ & $-12.5$\\
    10 & $-50.0008998\phantom{00}$ & $-50.00090875$ & $-50.0$\\
    50 & $-1250.55965\phantom{0000}$ & $-1250.697356\phantom{00}$ & $-1250.0$\\
    \bottomrule
  \end{tabular}
\end{table}

In order to validate the quality of the ground state energies obtained
by the Lanczos algorithm we also compared these to another numerical
solution, which was obtained using the Fourier-spectral method as
described for example in
\cite{Feit:Fleck:etal:1982:Solution_of,Bauke:Keitel:2011:Accelerating}.
Here the wave function is evaluated on a regular grid and propagated in
time via the Fourier split-operator method.  The bound-state energy
values follow from the peaks of the wave-function's auto-correlation
spectrum.  Table~\ref{tab:soft_core_eigenstates_Z_50} shows the energies
of the first nine bound states computed by both methods.  For the lowest
lying states agreement is excellent.  Energies of higher excited states
slightly disagree.  These states are broader than the ground state and,
therefore, boundary effects start to set in resulting in different
energy values for both methods.

Table~\ref{tab:soft_core_groundstates} shows the ground-state energies
of the two-dimen\-sional soft-core potential \eqref{eq:soft_core_pot} for
various values of $Z$ as obtained from the Lanczos algorithm with the
pseudospectral discretization and from first order perturbation theory
where lowest order relativistic corrections to the solution
\eqref{eq:SC_exact_gs} of the nonrelativistic Schr{\"o}dinger equation are
taken into account. For comparison also the nonrelativistic energies are
shown.  For small values of $Z$ the numerical values from the Lanczos
method are in excellent agreement (up to eight digits) with the
perturbation theory results.  The larger $Z$, however, the larger the
relativistic effects and for $Z=50$ the perturbation-theory result
differs from the numerical value as obtained by the Lanczos method by
about 0.1\,\textperthousand.

\subsection{Zeeman effect}

We consider the Coulomb potential with an external magnetic field of
magnitude $B$ as an application of the Lanczos algorithm to a fully
three-dimensional problem. The magnetic field leads to the well-known
Zeeman splitting of the degenerate hydrogenic eigenstates. In this
section, we analyze the splitting of the ground state in detail. A
perturbative solution of the relativistic splitting of the hydrogenic
ground state is known to be \cite{Mott:1971:Atomic_Collisions}
\begin{equation}
  \label{eq:dirac_ground_state_splitting}
  \Delta E = 
  \frac{B}{3 m} \left(1 + 2 \sqrt{1 - \left(\frac{Z}{c}\right)^2}\right) \, . 
\end{equation}

\begin{figure}
  \centering
  \includegraphics[width=0.95\linewidth]{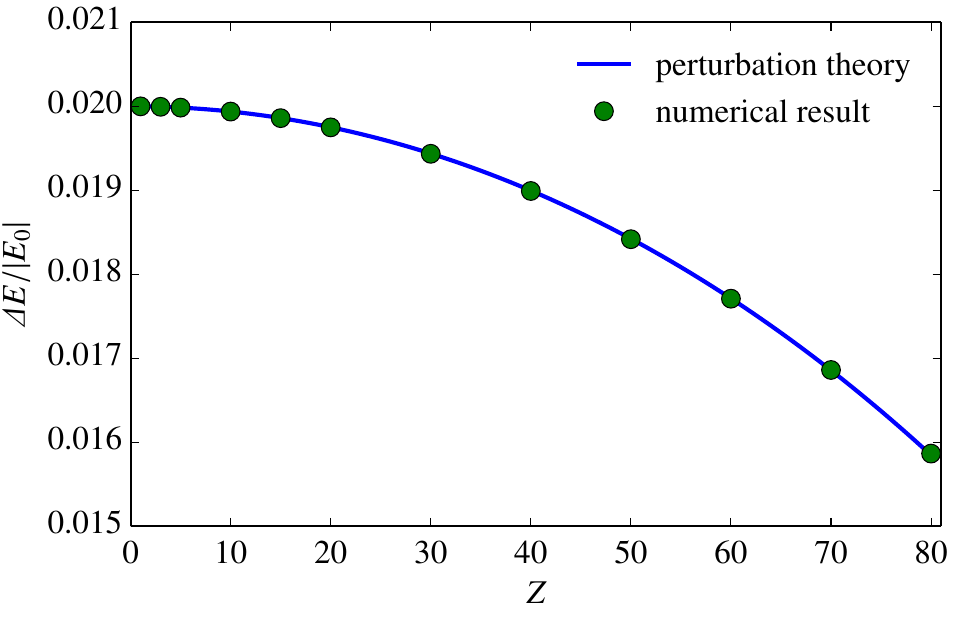}
  \caption{Relative Zeeman splitting of the Coulomb ground state of a
    hydrogen-like atom for a weak magnetic field $B=\num{0.01} \cdot
    Z^2$.  The numerical solution matches the perturbative solution.
    The system is discretized using a pseudospectral method with $N=128$
    basis functions per dimension and the Lanczos algorithm with 4000
    iterations.}
  \label{fig:zeeman}
\end{figure}

In Fig.~\ref{fig:zeeman} we analyze the relative Zeeman splitting
${\Delta E}/{|E_0|}$, where $E_0$ is the hydrogenic ground state energy
(minus the rest-mass energy $mc^2$) and $\Delta E$ the energy splitting
caused by the magnetic field $B$ that scales with the atomic number $Z$
as $B=\num{0.01} \cdot Z^2$.  Thus the magnetic field can be considered
as weak compared to the strength of the electric field close to the
atomic core and consequently the numerical results are in good agreement
with the perturbation-theory result
\eqref{eq:dirac_ground_state_splitting}.

\begin{figure}
  \centering
  \includegraphics[width=0.95\linewidth]{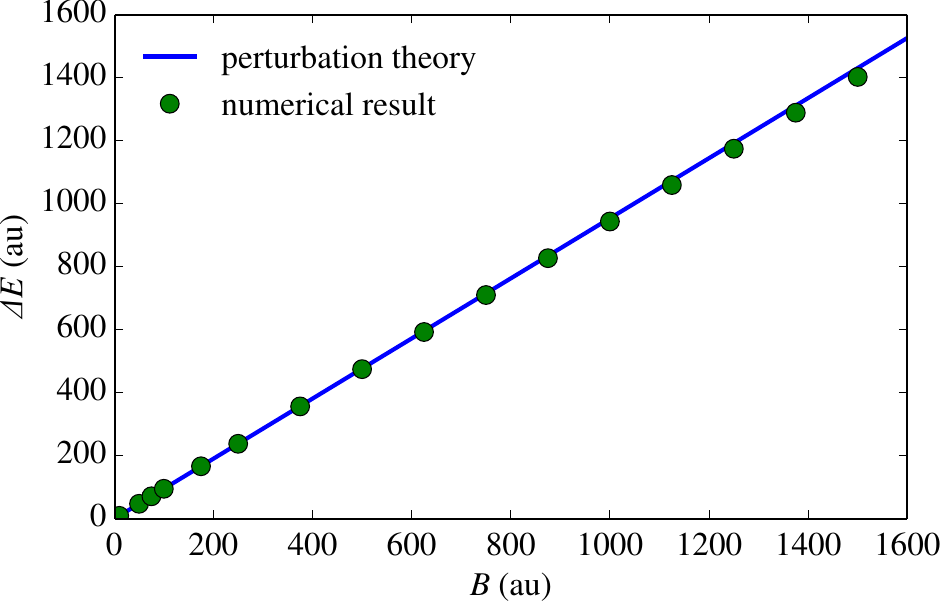}
  \caption{Difference of the relative ground-state splitting between the
    perturbative result \eqref{eq:dirac_ground_state_splitting} and our
    numerical result for a hydrogen-like atom with $Z = 50$ fixed. For
    increasing field-strength the numerical solution disagrees with the
    perturbative result as expected. For weak magnetic fields the
    agreement is excellent.}
  \label{fig:zeeman_z_const}
\end{figure}

Figure~\ref{fig:zeeman_z_const} shows the difference between our
numerical result and the perturbation-theory result
\eqref{eq:dirac_ground_state_splitting} for the fixed atomic number $Z =
50$ as a function of the magnetic field strength $B$.  For weak magnetic
fields, the numerical solution by the Lanczos method agrees very well
with perturbation theory.  Both results disagree, however, for fields
stronger than $B\approx\SI{1000}{\au}$ by a few percent.  Here we leave
the parameter domain where perturbation theory is applicable.

\section{Time-dependent problems}
\label{sec:time}

After demonstrating the application of the Lanczos algorithm to
time-independent relativistic eigenproblems we address time-dependent
problems now.  The motion of a free wave packet is one of the rare
systems where an analytical solution to the time-dependent Dirac
equation is known.  Thus, it provides an ideal benchmark system for
time-dependent problems.  We chose a relativistic one-dimensional wave
packet containing both positive-energy and negative-energy free-particle
states with Gaussian momentum distribution. When using the standard
position operator, the center of mass $\left<x \right> =
\braket{\Psi|x|\Psi}$ shows so-called Zitterbewegung, which is caused by
an interference between positive and negative energy contributions.

\begin{figure}
  \centering
  \includegraphics[width=0.95\linewidth]{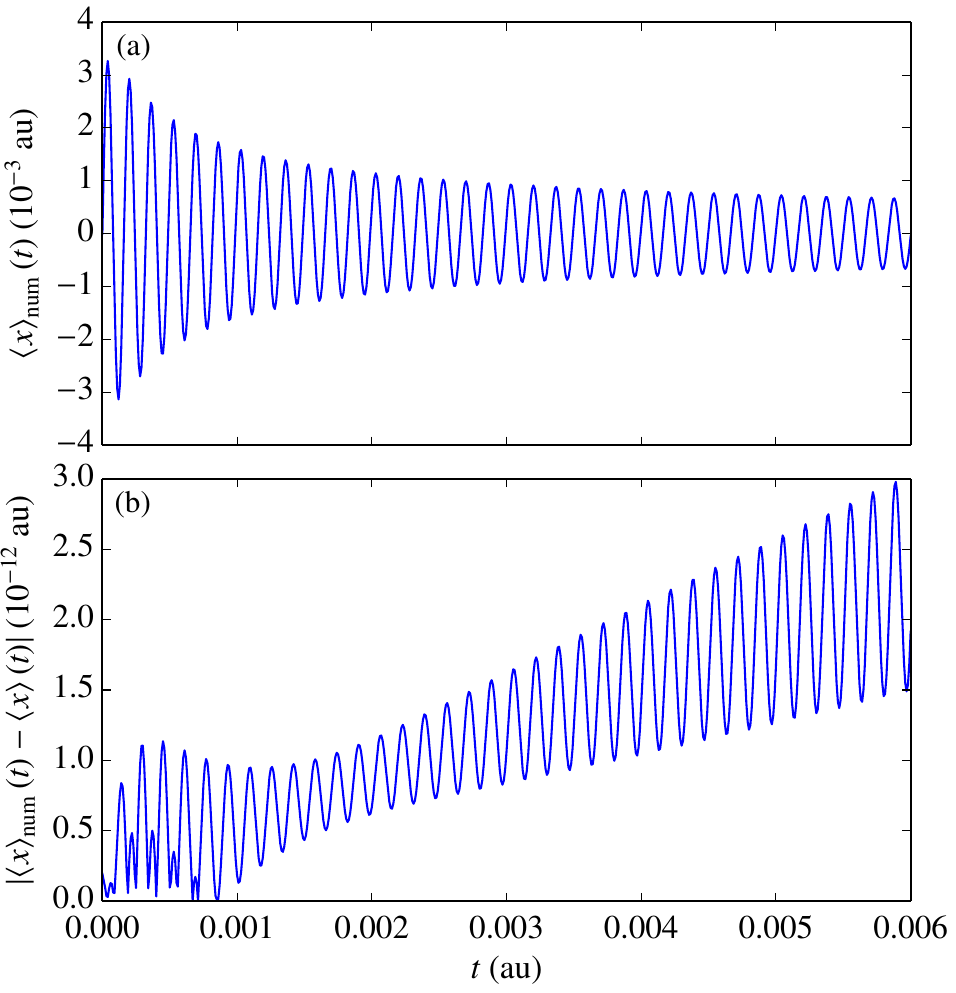}
  \caption{(a) Center-of-mass evolution of a highly relativistic wave
    packet with vanishing mean momentum. It was set up to contain both
    positive and negative energy components such that Zitterbewegung can
    be observed. (b) Difference between the numerical solution based on
    the Lanczos algorithm and the exact center-of-mass evolution. The
    numerical solution was obtained with $N = 512$ basis functions and
    $k=8$ Lanczos iterations per time step.}
  \label{fig:zitterbewegung_position}
\end{figure}

In our numerical test we employed the pseudospectral method with a basis
set of 512 basis functions for the spatial discretization of the wave
function.  Time steps of size $\Delta t = \SI{e-5}{\au}$ and $k = 8$
Lanczos iterations per time step were used.  The numerical result for
the wave function $\Psi_\mathrm{num}(x, t)$, which is represented by the
vector $\vec\psi(t)$, can be compared to the analytical solution
\begin{equation}
  \label{eq:Psi_zitter}
  \Psi(x, t) = 
  \frac{1}{\sqrt{2\pi\hbar}}
  \int
  \frac{\e^{-p^2/(4\sigma^2)}}{\sqrt[4\,]{\smash[b]{2\pi \sigma^2}}}
  \frac1{\sqrt{2}}
  \big( u^+(p) + u^-(p) \big)
  \,\D p\,,
\end{equation}
with the free-particle momentum states of positive and negative energy
\begin{subequations}
  \begin{align}
    u^+(p) & =
    \begin{pmatrix}
      d^+(p) \\ \sgn(p)d^-(p)
    \end{pmatrix} \e^{\I(px - t E(p))/\hbar} \,,\\
    u^-(p) & =
    \begin{pmatrix}
      -\sgn(p)d^-(p) \\ d^+(p)
    \end{pmatrix}\e^{\I(px + t E(p))/\hbar} \,,
  \end{align}
\end{subequations}
which are defined via
\begin{equation}
  d^\pm(p) =
  \left(\frac{1}{2} \pm \frac{1}{2\sqrt{1 + p^2/(mc)^2}}\right)^{1/2} 
\end{equation}
and 
\begin{equation}
  E(p) = \sqrt{ m^2c^4 + c^2p^2}\,.
\end{equation}
The width of the momentum distribution was chosen as
$\sigma=\SI{50}{\au}$.  The numerical result of the wave packet's
center-of-mass motion is presented in
Fig.~\ref{fig:zitterbewegung_position}(a).  It features the expected
oscillation around the origin with decaying amplitude for larger
times. Figure~\ref{fig:zitterbewegung_position}(b) shows the difference
$|\braket{x}_\mathrm{num}(t)-\braket{x}(t)|=|\braket{\Psi_\mathrm{num}(t)|x|\Psi_\mathrm{num}(t)}-\braket{\Psi(t)|x|\Psi(t)}|$.
On average the center-of-mass error increases linearly with time $t$ but
remains at the order of $\SI{e-12}{\au}$ which is eight orders of
magnitude smaller than the amplitude of the Zitterbewegung.

\begin{figure}
  \centering
  \includegraphics[width=0.95\linewidth]{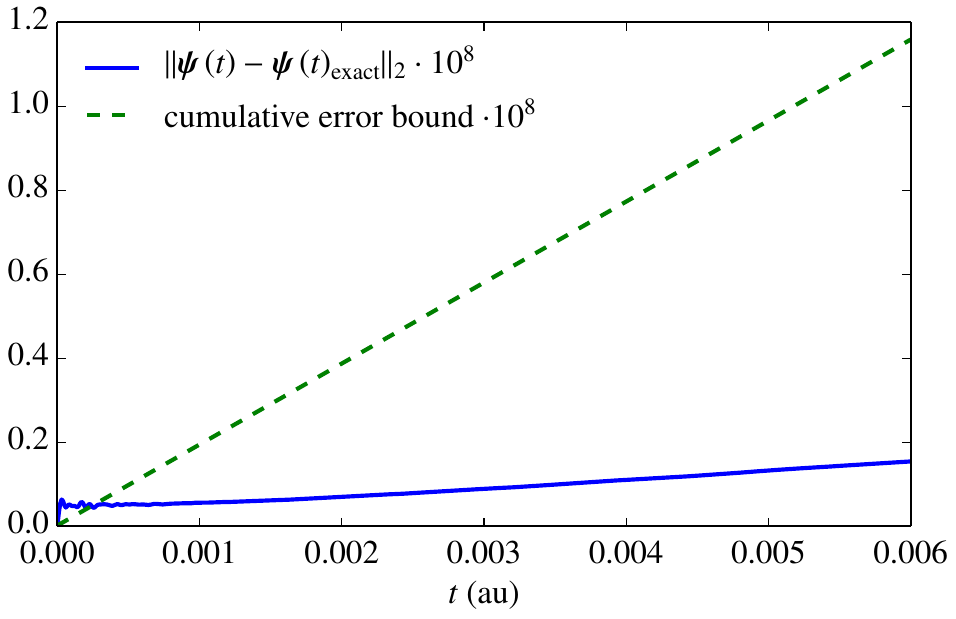}
  \caption{The cumulative error bound (error bound at each time step given
    by \eqref{eq:matrix_exponential_error}) and the actual numerical
    error as functions of the time for the simulation shown in
    Fig.~\ref{fig:zitterbewegung_position}.}
  \label{fig:zitterbewegung_error}
\end{figure}

Note that if the temporal step width is not too large then the change of
the wave function from time $t$ to $t+\Delta t$ will be small and
propagating the wave function via \eqref{eq:matrix_exp_approx} can be
quite accurate even for relatively small Krylov subspace dimensions.  An
upper bound of the error that is introduced due to the Lanczos
approximation of the matrix exponential can be estimated via
Eq.~\eqref{eq:matrix_exponential_error}. Thus, an upper bound of the
total error in a series of time steps can be estimated by adding up the
estimates \eqref{eq:matrix_exponential_error} at every time step. We
call this estimate the cumulative error bound, which is indicated by the
dashed line in Fig.~\ref{fig:zitterbewegung_error}. For the specific set
of parameters the error bound \eqref{eq:matrix_exponential_error} turns
out to be almost constant and therefore the cumulative error bound grows
linearly.  In addition, we computed at every time step the real error of
the wave function defined as
$\lVert\vec{\psi}(t)-\vec{\psi}_\mathrm{exact}(t)\rVert_2$, where the
vector $\vec{\psi}_\mathrm{exact}(t)$ equals the analytical solution
$\Psi(x, t)$ given by \eqref{eq:Psi_zitter} at the collocation points.
This error is indicated by the solid line in
Fig.~\ref{fig:zitterbewegung_error}.  Comparing the two lines in
Fig.~\ref{fig:zitterbewegung_error} illustrates that the error bound
\eqref{eq:matrix_exponential_error} indeed overestimates the actual
numerical error.  At the end of the propagation at $t=\SI{0.006}{\au}$
the error of the wave function is almost an order of magnitude smaller
than the cumulative estimate, despite the fact that the error introduced
by discretization is not included in the estimate
\eqref{eq:matrix_exponential_error}.  Note that, in
Fig.~\ref{fig:zitterbewegung_error} it appears as if for times
$t<\SI{0.0003}{\au}$ the numerical error
$\lVert\vec{\psi}(t)-\vec{\psi}_\mathrm{exact}(t)\rVert_2$ would be
larger than the cumulative error bound. This, however, is an artifact
caused by the numerical evaluation of the exact wave function
\eqref{eq:Psi_zitter} which involves a numerical integration via a
discrete Fourier transform on a regular grid and interpolation to the
collocation points.  In conclusion, our results show that very precise
approximations of the time evolution operator of the time-dependent
Dirac equation can be calculated by the Lanczos algorithm.

\section{Parallel implementations}
\label{sec:implementation}

Our evaluation of the Lanczos propagator for the Dirac equation was
mainly motivated by the fact that the Lanczos algorithm has the
potential to scale well in parallel implementations on various parallel
computing architectures including distributed memory systems.  Thus, it
might be suitable for large-scale computations.  A parallel Lanczos
algorithm only requires parallel computation of inner products and a
parallel implementation of the Hamiltonian's action on a state vector.

In the program Dirac\_Laczos we implemented a parallel version of the
Lanczos propagator for the two-dimensional Dirac equation by utilizing
the Message Passing Interface (MPI) standard \cite{MPI3.0:2012}.  The
differential operators of the Hamiltonian were approximated via first
order finite differences.  In this case (and in contrast to the
pseudospectral method used in the previous sections) the differentiation
matrices are sparse and the Hamiltonian's action on a state vector can
be parallelized efficiently via decomposition of the total rectangular
computational grid into smaller sub-grids; one sub-grid per process.
This domain decomposition approach can also be applied to other
finite-difference based algorithms, for example, real space split
operator schemes for the Klein-Gordon equation
\cite{Ruf:2009:Klein-Gordon_Splitoperator} and the Dirac equation
\cite{Fillion-Gourdeau:Lorin:etal:2012:Numerical_solution_time-dependent}.
The exchange of boundary data between neighboring domains is implemented
via the nonblocking MPI functions \code{MPI_Isend} and \code{MPI_Irecv}
\cite{Bauke:Mertens:2005:ClusterComputing}.  Parallel scalar products
are calculated via \code{MPI_Allreduce}.

\begin{figure}
  \centering
  \includegraphics[width=\linewidth]{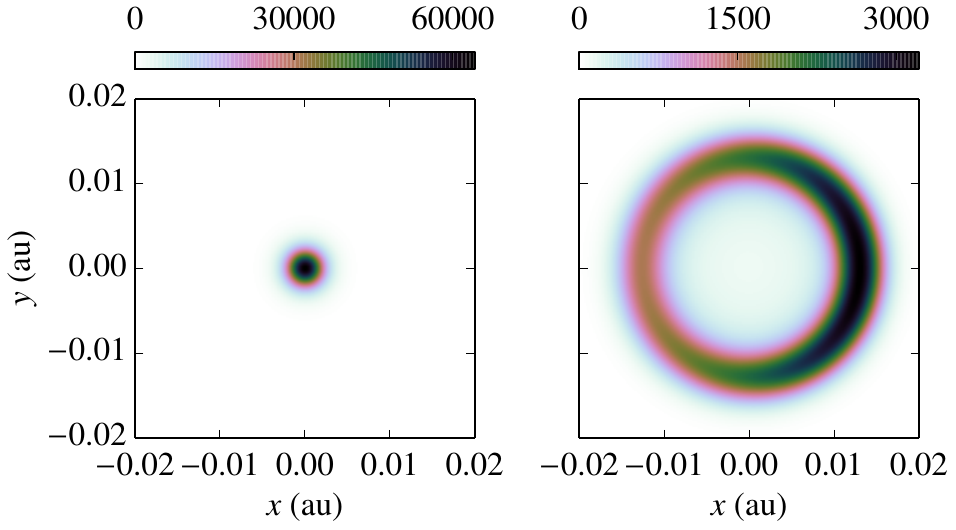}
  \caption{Two-dimensional wave packet containing a Gaussian
    superposition of positive-energy free-particle states with a broad
    momentum distribution. The left part shows the density $|\Psi(\vec
    r, t)|^2$ at time $t=\SI{0}{\au}$ and the right part at
    $t=\SI{0.0001}{\au}$.}
  \label{fig:wave_packet_2d}
\end{figure}

For the following benchmarks we consider a free two-dimen\-sional wave
packet. It is constructed by a Gaussian superposition of positive-energy
free-particle states with mean momentum $\vec p=(\SI{100}{\au},
\SI{0}{\au})$ and a momentum space width of $\sigma_x = \sigma_y =
\SI{400}{\au}$. Due to the broad momentum distribution and the nonlinear
relativistic relation between velocity and momentum a shock front with a
ring structure emerges during the temporal evolution of the wave packet
as shown in Fig.~\ref{fig:wave_packet_2d}. The figure's left part shows
the initial wave packet and the right part shows the wave packet at time
$t=\SI{0.0001}{\au}$. The nonzero mean momentum in $x$~direction leads
to a motion of the wave-packet's center-of-mass into this direction.
The calculation was performed on a regular grid of $8192 \times 8192$
grid points and employing $k=10$ Lanczos iterations per time step of
size $\Delta t=\SI{e-7}{\au}$ each.  Due to the rather large grid that
was used here the sequential propagation of a single time step took
about two minutes on the employed hardware.

\begin{figure}
  \centering
  \includegraphics[width=0.95\linewidth]{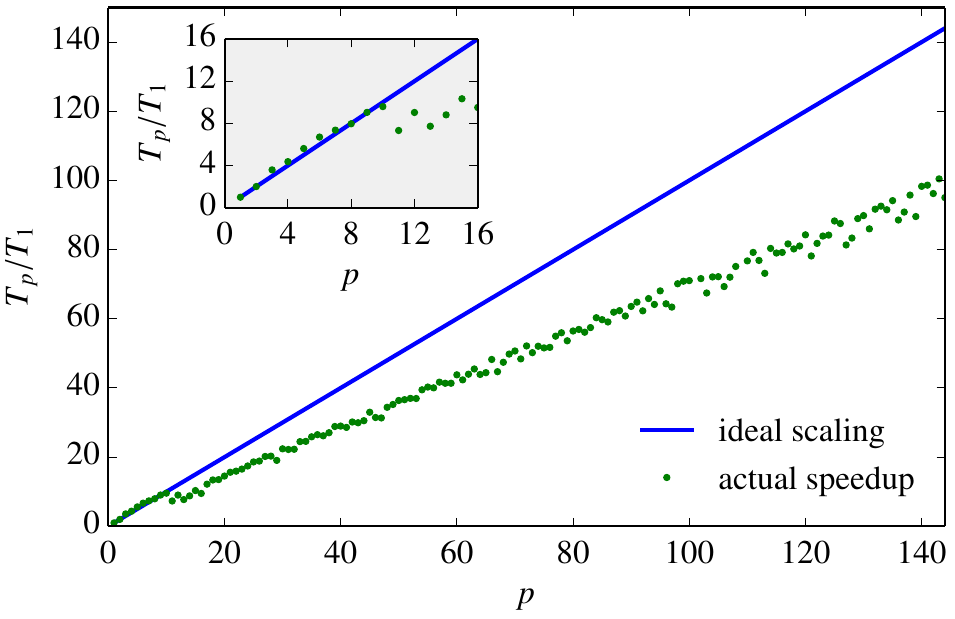}
  \caption{Speedup of the Dirac Lanczos propagator in our MPI
    implementation as a function of the number of parallel processes
    $p$.}
  \label{fig:speedup}
\end{figure}

To analyze the scaling properties of the Dirac Lanczos propagator, the
propagation of the wave packet in Fig.~\ref{fig:wave_packet_2d} was
carried out with a varying number of parallel processes $p$ giving the
run time $T_p$.  Comparing this to a sequential calculation yields the
speedup $s_p=T_1/T_p$.  The calculations were performed on a homogeneous
cluster of nodes with dual Xeon-E5 CPUs with 8 cores each (16 cores per
node).  The nodes were connected with a 10~Gigabit Ethernet
interconnect.  With increasing number of processes the nodes were filled
up successively.  In order to avoid performance degeneration due to
intra-node process migration between different CPUs each process was
pinned to a specific core.  The measured speedup is shown in
Fig.~\ref{fig:speedup}.  The speedup is approximately linear in the
number of processes but differs from the ideal scaling $s_p\approx p$
due to the communication overhead, which also grows with the number of
processes.

The discretization by finite differences leads to a consecutive
memory-accesses pattern when calculating the inner products and applying
the Hamiltonian with very few accesses per fetched memory element.
Consequently, the CPUs' cache hierarchy cannot be used efficiently, so
that the memory bandwidth limits the performance.  The inset of
Fig.~\ref{fig:speedup} illustrates the memory-bandwidth limitation of
the Lanczos propagator. For $p\le 10$, that is when a node with 16 cores
is still under-utilized, we have an almost ideal scaling with
$s_p\approx p$. But for $p>10$ limitations due to the memory bandwidth
set in.  This efficiency degradation occurs before the first node is
completely filled with 16 processes.  Thus, network traffic is not the
main performance limiting factor.  Further numerical experiments showed
that the speedup and the parallel efficiency can be increased when the
nodes are systematically under-utilized. This means that, for example,
only four cores per CPU are used.  This also supports our assertion that
the Lanczos propagator is memory-bandwidth bounded.

\section{Conclusions}

We evaluated the Lanczos algorithm for the application to the
time-independent and the time-dependent Dirac equation. Our results
indicate that the Lanczos algorithm is able to compute precise solutions
of the Dirac equation.  This was demonstrated for the time-independent
eigenvalue problem solving the two-dimensional soft-core potential and
the three-dimensional Coulomb potential with an additional strong
magnetic field.  While in the case of the soft-core potential the
numerical instabilities of the Lanczos algorithm were circumvented by
applying full reorthogonalization, the plain Lanczos algorithm was
applied to the Coulomb potential with magnetic field. As in the case of
the soft-core potential the Lanczos approach allowed us to calculate the
pair of bound states with lowest energy, such that we were able to
calculate the Zeeman splitting in a strong magnetic field.  Furthermore,
we demonstrated for one- and two-dimensional wave packets that the
Dirac-equation's time evolution operator can be approximated very
precisely using the Lanczos algorithm. 

The Lanczos approach is not specific to particular means of
discretization of the Dirac Hamiltonian as long as the discretization
preserves Hermiticity.  Here we employed pseudospectral methods and
finite differences.  The latter yields a sparse matrix representation of
the Dirac Hamiltonian and in this case the Lanczos propagator can be
parallelized very efficiently via domain decomposition as demonstrated
in this paper. The benefit of pseudospectral discretizations is to
approximate the differential operators of the Dirac Hamiltonian very
precisely and in this way allowing for very accurate bound-state
calculations and time-propagation.  The accuracy of the Lanczos Dirac
propagator combined with its property to perform well on modern parallel
hardware architectures makes it a good choice for large-scale
calculations of relativistic quantum dynamics.  A specific
implementation of the Lanczos propagator for the Dirac equation is given
in the program Dirac\_Laczos, which utilizes message passing for
parallelization.  It might be interesting to implement the Dirac Lanczos
propagator also on other parallel hardware architectures, as for example
graphics processing units.

\secreferences
\bibliography{Krylov_Dirac}

\end{document}